\documentclass[aps,prl,amsmath, twocolumn, groupedaddress, showpacs]{revtex4}
\usepackage[T1]{fontenc}
\usepackage[latin9]{inputenc}
\usepackage{amstext}
\usepackage{graphicx}
\usepackage{amssymb}

\makeatletter

\providecommand{\LyX}{L\kern-.1667em\lower.25em\hbox{Y}\kern-.125emX\@}
\providecommand{\tabularnewline}{\\}

\makeatother

\begin{document}

\preprint{}

\title{Energy decay and frequency shift of a superconducting qubit \\ from non-equilibrium quasiparticles}

\author{ M. Lenander$^1$}
\author{ H. Wang$^{1,2}$}
\author{ Radoslaw C. Bialczak$^1$}
\author{ Erik Lucero$^1$}
\author{ Matteo Mariantoni$^1$}
\author{ M. Neeley$^1$}
\author{ A. D. O'Connell$^1$}
\author{ D. Sank$^1$}
\author{ M. Weides$^1$}
\author{ J. Wenner$^1$}
\author{ T. Yamamoto$^{1,3}$}
\author{ Y. Yin$^1$}
\author{ J. Zhao$^1$}
\author{ A. N. Cleland$^1$}
\author{ John M. Martinis$^1$}

\affiliation{$^1$Department of Physics, University of California, Santa Barbara, CA 93106, USA}
\affiliation{$^2$Department of Physics, Zhejiang University, Hangzhou 310027, China}
\affiliation{$^3$Green Innovation Research Laboratories, NEC Corporation, Tsukuba, Ibaraki 305-8501, Japan}

\begin{abstract}
Quasiparticles are an important decoherence mechanism in superconducting qubits, and can be described with a complex admittance that is a generalization of the Mattis-Bardeen theory.  By injecting non-equilibrium quasiparticles with a tunnel junction, we verify qualitatively the expected change of the decay rate and frequency in a phase qubit.  With their \textit{relative} change in agreement to within 4\,\% of prediction, the theory can be reliably used to infer quasiparticle density.  We describe how settling of the decay rate may allow determination of whether qubit energy relaxation is limited by non-equilibrium quasiparticles.
\end{abstract}


\maketitle

Superconducting resonators and Josephson qubits are remarkable systems
for quantum information processing \cite{Clarke08}, with recent experiments demonstrating 3-qubit entanglement \cite{Neeley10, DiCarlo10}, Bell and Leggett-Garg inequalities \cite{Ansmann09, Saclay10}, resonance fluorescence \cite{Astafiev10}, photon jumps \cite{Vijay10} and transducers \cite{shell:game}, and NOON states \cite{NOON10}.   For maximum coherence, devices are operated at temperatures $T$ well below the superconducting transition $T_{c}$ to suppress thermal excitations and quasiparticles.  Nevertheless, experiments using resonators and qubits can sense the presence of quasiparticles, which arise from non-equilibrium sources such as stray infrared radiation, cosmic rays, or energy relaxation in materials \cite{resqps,Martinis09}.  It is vitally important to completely understand the theory of quasiparticle dissipation, determine whether it limits qubit coherence, and identify generation mechanisms.

In this Letter, we describe the effect of non-equilibrium quasiparticles with linear response theory \cite{Yalethy} including dissipative and dispersive terms, and then experimentally test that theory by measuring the response of a superconducting phase qubit to the injection of non-equilibrium quasiparticles.  The decay rate of the qubit is shown qualitatively to depend on the injection and recombination rate of quasiparticles.  Because the density of quasiparticles $n_\textrm{qp}$ is not known from injection, we compare the decay rate and frequency shift of the qubit with changing $n_\textrm{qp}$ to test quantitatively the correct dependence with theory.  We also show that the characteristic settling time to steady state is consistent with theory, and  may allow an additional measure of quasiparticle density.

In superconductivity, non-equilibrium quasiparticles at energy $E$ can be described using an occupation probability $f(E)$.  Quasiparticle effects can be expressed in a two-fluid manner by considering their total density
$n_\textrm{qp} = 2D(E_f)\int_\Delta^\infty \rho(E) f(E) dE$,
where $D(E_f)/2$ is the single-spin density of states, $\Delta$ is the gap, and $\rho(E)=E/\sqrt{E^2-\Delta^2}$ is the normalized density of quasiparticle states.  For low dissipation appropriate for qubits, we need only consider small occupations $n_\textrm{qp} \ll n_\textrm{cp}$, where we define $n_\textrm{cp} \equiv D(E_f) \Delta$ as the density of Cooper pairs.  Using BCS theory, non-equilibrium quasiparticles lower the superconducting gap as $\Delta = \Delta_0(1-n_\textrm{qp}/n_\textrm{cp})$, where $\Delta_0$ is the gap without quasiparticles \cite{supp}.

We first describe the effect of non-equilibrium quasiparticles on Josephson tunneling.  For arbitrary occupation $f(E)$, the Josephson inductance $L_J$ and admittance $Y_J$ from Cooper pair tunneling has frequency dependence
\begin{align}
Y_J(\omega) = \frac{1}{i\omega L_J } & =
\frac{1}{i\omega}\frac{2 \pi I_0 \cos\phi  }{\Phi_0 } [1-2f(\Delta)] \ ,
\label{eq:ZJ}
\end{align}
where $\phi$ is the junction phase difference, $\Phi_0 = h/2e$ is the flux quantum, and $I_0$ is the critical current \cite{supp}.  Non-equilibrium quasiparticles effectively lower the critical current by a factor $1-2f(\Delta)$ by blocking pair channels.  Note this reduction is not a function of $n_\textrm{qp}$, but depends on the occupation probability at the gap $f(\Delta)$.  For an exact calculation, the occupation would be for Andreev \textit{bound} states in the junction at energies slightly below the gap \cite{ABS}; for small tunneling probability, we assume this equals the occupation of \textit{free} quasiparticles at the gap $f(\Delta)$.  Thermal quasiparticles have an occupation $f_T(E)=1/[1+\exp(E/kT)]$, which yields the expected temperature dependence $\tanh(\Delta/2kT)$.

Defining $a$ through the relation $f(\Delta)=a\, n_\textrm{qp}/n_\textrm{cp}$, we find $a \simeq \sqrt{\Delta/2\pi kT}$ for thermal quasiparticles at low temperature \cite{ftherm}.  For the case of non-equilibrium quasiparticles generated by an external pair-breaking source, we numerically compute $a \simeq 0.12 \, (n_\textrm{qp}/n_\textrm{cp})^{-0.173}$, giving  $a \simeq 1.2$ for typical parameters \cite{supp}.

The total junction admittance $Y_j$ is also affected by quasiparticle tunneling \cite{Yalethy}, and is given by
\begin{align}
\frac{Y_j(\omega)}{1/R_n}=
&  \frac{\Delta}{\hbar\omega } (1+\cos\phi)
\left[ \frac{1+i}{\sqrt{2}} \sqrt{\frac{\Delta}{\hbar\omega} }
\frac{n_\textrm{qp}}{n_\textrm{cp}} - i \pi f(\Delta) \right] \nonumber \\
& -i \pi \frac{\Delta}{\hbar\omega}  \cos\phi \ [1-2f(\Delta)]
\label{eq:Zj} \ ,
\end{align}
where the last term, from Cooper pair tunneling, has the critical current $I_0=\pi\Delta/2eR_n$ re-expressed using the gap and junction resistance $R_n$. The first term gives a different phase dependence $1+\cos\phi$, and has contribution from the tunneling of free quasiparticles [$n_\textrm{q}$] and bound Andreev states [$f(\Delta)$].

For a junction phase difference $\phi=0$ the factors from $f(\Delta)$ cancel out, and the admittance is
\begin{align}
\frac{Y_j(\omega)}{1/R_n}  & = (1+i) \sqrt{2}\left(\frac{\Delta}{\hbar\omega}\right)^{3/2}
\frac{n_{\mathrm{qp}}}{n_{\mathrm{cp}}}-i\pi\frac{\Delta}{\hbar\omega} \ .
\label{eq:M-B}
\end{align}
This result is equivalent to the normalized conductivity $\sigma(\omega) / \sigma_{n}$ given by the Mattis-Bardeen theory \cite{MB} for thermal quasiparticles \cite{MBints} in the limit $kT \ll \hbar \omega \ll \Delta $.  This shows that resistance in a superconductor can be modeled as a series of Josephson junctions.

Quasiparticle damping increases the energy relaxation rate in Josephson qubits.  For the phase qubit, where matrix elements are well approximated by harmonic oscillator values, the relaxation rate is given by \cite{Esteve86}
\begin{align}
\Gamma_1 \simeq \frac{\textrm{Re}\{Y_j(E_{10}/\hbar) \} }{C}
=  \frac{1+\cos\phi}{\sqrt{2} R_n C} \left( \frac{\Delta}{E_{10}}\right)^{3/2}
\frac{n_\textrm{qp}}{n_\textrm{cp}} \ , \label{eq:gamma1}
\end{align}
where $C$ is the junction capacitance and $E_{10}$ is the qubit energy.  This result is identical to that found from environmental $P(E)$ theory in the low impedance limit \cite{Martinis09}, and now includes a $\cos\phi$ term coming from the interference of electron- and hole-like tunneling events \cite{Frankthy}.

Quasiparticles also change the imaginary part of $Y_j$, which then shifts the qubit frequency by $\delta E_{10}/\hbar \simeq -\textrm{Im}\{Y_j(E_{10}/\hbar) \} / 2C $ using perturbation theory \cite{Esteve86}.  For the $n_\textrm{qp}/n_\textrm{cp}$ term in quasiparticle tunneling, the real and imaginary parts of $Y_j$ are equal, which yields a change in angular frequency $-\Gamma_1/2$.  Quasiparticles reduce the admittance from  Cooper pair tunneling by the factor $1-(1+2a)n_\textrm{qp}/n_\textrm{cp}$ due to a change in $\Delta$ and the pair-blocking terms.  Including the final $f(\Delta)$ term from quasiparticle tunneling, the frequency shift is
\begin{align}
\frac{\delta E_{10}'}{h\Gamma_1}=
-\frac{1}{4\pi}
\left[ 1 - \frac{1}{b}
\, \frac{a-(1+a)\cos\phi}{1+\cos\phi} \right]
\label{eq:slope}
\end{align}
where $b = \sqrt{\Delta/2E_{10}}/\pi \simeq 0.6$ for typical parameters.  Here, the shift is normalized to the damping since both are proportional to $n_\textrm{qp}$.

The above calculation assumes constant $\phi$, which is not valid for a phase qubit where we impose the constraint of constant current bias $I$.  Assuming the standard Josephson current-phase relationship, the critical current changes both from the gap and quasiparticle excitations in the junction, giving $\delta I_0/I_0 = -(1+2a)\, n_\textrm{qp}/n_\textrm{cp}$.  Because the qubit frequency scales as $E_{10} \propto (I_0-I)^{1/4}$, a change in qubit critical current changes the bias $I_0-I$ and results in an additional shift in frequency given by $\delta E_{10}/E_{10} = (1/4) \delta I_0/(I_0-I)$.

Since the equations for both dissipation and frequency shift are proportional to $n_\textrm{qp}$, their ratio can be related using qubit parameters.  Including both frequency shifts, we find \cite{supp}
\begin{align}
\label{eq:slopeFull}
\frac{\delta E_{10}}{h\Gamma_1} = \frac{\delta E_{10}'}{h\Gamma_1}
-\frac{1}{4} \frac{1+2a}{1+\cos\phi}
\left(\frac{I_0 }{I_0-I}\frac{E_{10}}{\Delta_0}\right)^{1/2} \ .
\end{align}
The latter term dominates because the current bias is typically set close to the critical current $I_0-I \ll I_0$.

\begin{figure}[b]
\includegraphics[scale=1.0]{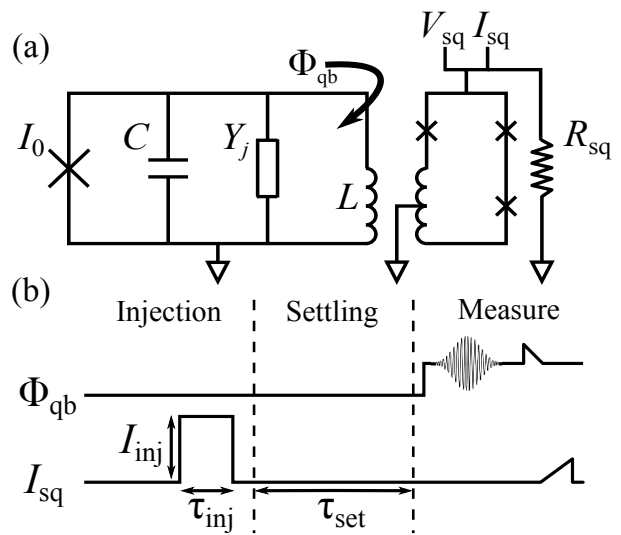}
\caption{\label{fig:schematic} (a) Schematic of phase qubit, with critical current $I_0 \simeq 2.0\,\mu\textrm{A}$ and capacitance $C \simeq 1.0\,\textrm{pF}$, shunted by admittance $Y_j$ coming from non-equilibrium quasiparticles.  The measurement SQUID, shunted by a resistor $R_\textrm{sq} = 30\,\Omega$, generates quasiparticles when switched into the voltage state. (b) Time sequence of experiment, showing initial pulse for quasiparticle injection at the SQUID, settling time $\tau_\textrm{set}$, flux pulses applied to the qubit for frequency or decay time measurement, and a final flux measurement by the SQUID. }
\end{figure}

As shown in Fig.\,\ref{fig:schematic}(a), we experimentally test this theory by using a phase qubit \cite{phaseqbdetail} to measure the effect of non-equilibrium quasiparticles injected via a separate tunnel junction in the SQUID.  The effect of quasiparticles are modeled by a parallel admittance $Y_j$ that changes the qubit $|1\rangle$ to $|0\rangle$ state transition frequency and decay rate.  Qubit measurement produces a state-dependent change in the loop flux that is measured with a SQUID readout circuit \cite{phaseqbdetail}.  The SQUID is also used to generate quasiparticles when driven into the voltage state $V_\textrm{sq} > 0$.  With the SQUID shunted by a resistor $R_\textrm{sq} = 30\,\Omega$, the SQUID current $I_\textrm{sq}$ can be adjusted to produce a voltage $V_\textrm{sq}$ from $\sim 0.6\, \Delta/e$ to above $2\Delta/e$, greatly changing the generation rate of quasiparticles.

The experimental time sequence is illustrated in Fig.\,\ref{fig:schematic}(b). Quasiparticles are initially generated by applying SQUID current to produce $V_\textrm{sq} \gtrsim 2\Delta/e$ for a time $\tau_\textrm{inj}$: quasiparticles then diffuse to the qubit via the ground plane connection.  The injection is turned off for a settling time $\tau_\textrm{set}$, after which we apply a flux waveform to perform a measurement of the qubit resonance frequency or decay rate \cite{phaseqbdetail}.  We then ramp $I_\textrm{sq}$ to measure the flux state for qubit readout, but at a low voltage $V_\textrm{sq} \simeq 0.6\,\Delta/e$ so that few quasiparticles are generated.  The experiment is typically repeated $\sim 1000$ times to produce probabilities.

\begin{figure}[t]
\includegraphics[scale=1]{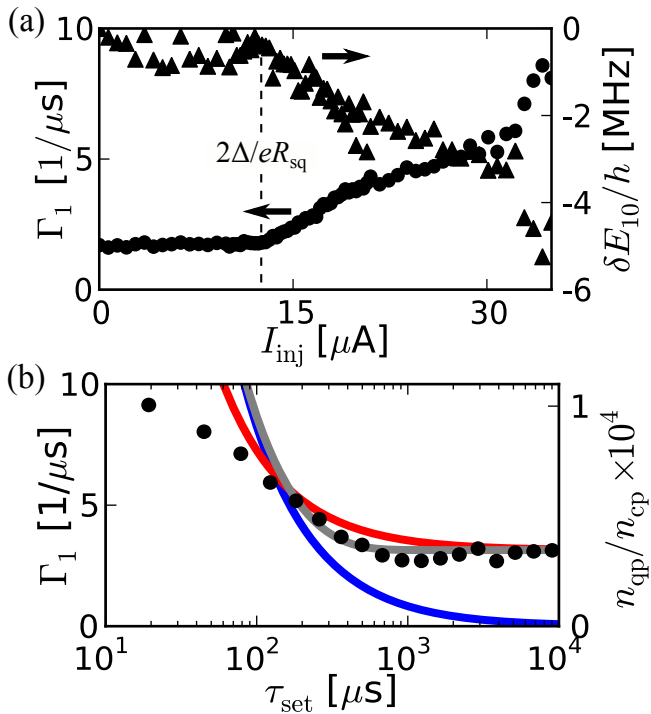}
\caption{\label{fig:settling}
Effect of quasiparticles on the phase qubit. (a) Plot of qubit decay rate $\Gamma_1$ (circles) and frequency shift $\delta E_{10}/h$ (triangles) as a function of SQUID injection current $I_\textrm{inj}$, for a fixed time of injection $\tau_\textrm{inj} = 200\,\mu\textrm{s}$ and settling $\tau_\textrm{set}= 300\,\mu\textrm{s}$.  Dashed vertical line indicates SQUID voltage $V_\textrm{sq} = 2\Delta/e$, above which quasiparticle generation increases rapidly.  (b) Plot of decay rate versus settling time (circles) for fixed injection current $I_\textrm{inj} = 20\,\mu\textrm{A}$ and time $\tau_\textrm{inj}= 200\,\mu\textrm{s}$.  Quasiparticles are observed to recombine on a time scale $\sim 300\,\mu\textrm{s}$. Bottom blue line is theory of Eq.\,(\ref{eqr0}) with electron-phonon coupling \cite{Kaplan} $\tau_0=400\,\textrm{ns}$.  Theories with additional qubit decay (top red, $\tau_0 = 200\,\textrm{ns}$) and non-equilibrium quasiparticles of Eq.\,(\ref{eqr1}) (gray,  $\tau_0=400\,\textrm{ns}$) are also shown, each fit with $\tau_0$ and a parameter matching $\Gamma_1$ at 10\,ms.  Qubit number is 2, with $I_0=2.03\,\mu\textrm{A}$ and $C = 1.063\,\textrm{pF}$. }
\end{figure}

We plot in Fig.\,\ref{fig:settling}(a) the decay rate and frequency shift of the qubit versus injected SQUID current for a fixed settling time.  When the current $I_\textrm{sq}$ produces a SQUID voltage $V_\textrm{sq} < 2\Delta/e$ (left of dashed vertical line), we observed almost no frequency shift $\delta E_{10}/h$ or change in the decay rate $\Gamma_1$.   However, when the voltage exceeds the gap and greatly increases the quasiparticle generation rate from pairbreaking, we observe a large increase in qubit decay rate and decrease in frequency.

The recombination of quasiparticles is tested by keeping the injection current and time constant, while varying the settling time, as shown in Fig.\,\ref{fig:settling}(b).  For short times corresponding to the highest quasiparticle densities, we see more rapid decay of the qubit.  For settling times greater than $\sim 500\,\mu\textrm{s}$ the decay rate approaches that without injection of quasiparticles.

Although these data show qualitative agreement with expectations, direct quantitative analysis is difficult because of uncertainties in predicting $n_\textrm{qp}$ from modeling the generation, recombination, and diffusion of the quasiparticles.  However, an accurate quantitative test of the theory can be obtained by comparing the change in decay rate with the change in qubit frequency, which are both expected to scale as $n_\textrm{qp}$.

\begin{figure}[t]
\includegraphics[scale=1.0]{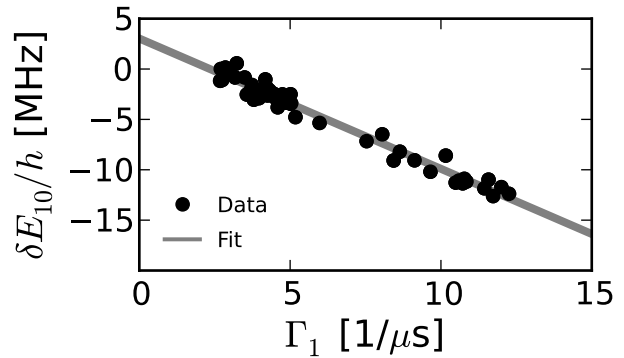}
\caption{\label{fig:qubit} Parametric plot of qubit frequency shift $\delta E_{10}/h$ and decay rate $\Gamma_1$ for various injection currents.  A linear relation is observed, consistent with theoretical predictions that both should scale as $n_\textrm{qp}$. Gray line indicates fit to the data.  }
\end{figure}

This comparison is shown in Fig.\,\ref{fig:qubit} for the range of injection currents plotted in Fig.\,\ref{fig:settling}.  With both quantities proportional to $n_\textrm{qp}$, a linear relation is expected and observed in the data.  The slope $\delta E_{10}/h \Gamma_1$ is thus a quantitative test of the theoretical prediction given in Eq.\,(\ref{eq:slopeFull}).  We extracted the slope from this type of plot for 3 devices and 5 experiments, as summarized in Table I. We find that the experimentally measured slope is on average only slightly larger ($3.6\%$) than predicted by theory and well within  experimental uncertainty (14\%).

We have also compared the dissipation and frequency shift in superconducting coplanar resonators made from aluminum, where quasiparticles are generated by raising the temperature.   We find the shift in resonance frequency and inverse quality factor also scale together, with a slope that is 0.77 times the predicted value \cite{supp}. The difference is probably due to two-level states \cite{Gao08}.

\begin{table}[t]
\caption{\label{tab:Slope Table} Qubit parameters showing average $\cos\phi$, bias parameter $(I_0-I)/I_0$, operating frequency,  slope extracted from experiment data as plotted in Fig.\,\ref{fig:qubit}, slope predicted by theory of Eq.\,(\ref{eq:slopeFull}), and their ratio. We observe experimental agreement within experimental uncertainty, as denoted in parenthesis.  Qubits 3, 4, and 5 are the same device tested at different times. Calculation of $\cos\phi$ and $(I_0-I)/I_0$ accounts for the inductive shunt in the flux-biased phase qubit.}
\begin{tabular}{ccccccc}
\hline
\hline
\ \ \ \  & $\cos\phi$ & $\frac{I_{0}-I}{I_{0}}$ & $ \frac{E_{10}}{h}$(GHz)  &
$[\frac{\delta E_{10}}{h\Gamma_1}]_\textrm{Exp}$ &
$[\frac{\delta E_{10}}{h\Gamma_1}]_\textrm{Thy}$ &
$\frac{\textrm{Exp}}{\textrm{Thy}}$\tabularnewline
\hline
1 & 0.12(3) & 0.042(5) & 6.523 & -1.42(9) & -1.34(19)& 1.05(16) \tabularnewline

2 & 0.17(4) & 0.066(6) & 6.743 & -1.30(3) & -1.03(13) & 1.22(15) \tabularnewline

3 & 0.19(4) & 0.071(3) & 6.413 & -1.13(6) & -1.03(09) & 1.09(12) \tabularnewline

4 & 0.33(5) & 0.130(3)  & 7.375  &-0.67(2)& -0.80(14)& 0.83(14) \tabularnewline

5 & 0.19(4) & 0.071(4) & 6.413  & -1.04(6) & -1.04(10) & 0.99(11) \tabularnewline
\hline
\hline
\end{tabular}
\end{table}

With confidence that we can accurately extract $n_\textrm{qp}$ from our experiment, we now analyze the time dependence of the data plotted in Fig.\,\ref{fig:settling}(b).  The decay of quasiparticles can readily be calculated for the case of small density when they mostly have energy near the gap \cite{supp}.  In this limit, the integral of Eq.\,(7) in Ref.\,\cite{Martinis09} gives a recombination rate $\Gamma_r=(21.8/\tau_0) (n_\textrm{qp}/n_\textrm{cp})$, where $\tau_0 \simeq 400\,\textrm{ns}$ is the characteristic electron-phonon coupling for aluminum \cite{Kaplan}. If $r_\textrm{qp}$ is the injection rate, the differential equation $dn_\textrm{qp}/dt = - 2 \Gamma^r n_\textrm{qp} + r_\textrm{qp}$ can be solved to give
\begin{align}
\frac{n_\textrm{qp}}{n_\textrm{cp}} &=  \frac{\tau_0/43.6}{t - t_0} \ \ \ \ \ \ \  \textrm{(case } r_\textrm{qp} = 0 ) \ , \label{eqr0} \\
n_\textrm{qp} &=  (n_\textrm{qp})^\textrm{eq} \coth[\,(\Gamma^r)^\textrm{eq} (t-t_0) \, ] \ , \label{eqr1}
\end{align}
with equilibrium values $(\Gamma^r)^\textrm{eq} = [(43.6/\tau_0)(r_\textrm{qp}/ n_\textrm{cp})]^{1/2}$ and $(n_\textrm{qp})^\textrm{eq}/n_\textrm{cp} = (\tau_0/43.6) (\Gamma^r)^\textrm{eq}$
, where $t_0$ is the integration constant.  After quasiparticle injection, Eq.\,(\ref{eqr1}) describes a decay that initially follows Eq.\,(\ref{eqr0}), but levels off to a steady state value $(n_\textrm{qp})^\textrm{eq}$ after time $\simeq 1/(\Gamma^r)^\textrm{eq}$.

We plot in Fig.\,\ref{fig:settling}(b) the prediction for the case of no equilibrium quasiparticles $r_\textrm{qp} = 0$ [Eqs.\,(\ref{eq:gamma1}) and (\ref{eqr0})], which depends only on $\tau_0$.
To account for the constant rate at times $\gtrsim 1\,\textrm{ms}$, we consider two hypotheses.  One is the qubit has an additional dissipation mechanism, so that the net decay rate is the sum of the above prediction with a constant offset.  The second assumes constant injection of non-equilibrium quasiparticles, as given by Eq.\,(\ref{eqr1}).  We plot these two additional predictions, which show somewhat different dependencies with time.

For settling times $\lesssim 300\,\mu\textrm{s}$, we believe the departure from theory is due to diffusion effects that are not included in this simple model \cite{Martinis09}.  Numerical simulations \cite{supp} show such deviations are reasonable given that the injection and qubit junctions are well separated.  Note the small dip in $\Gamma_1$ at time $\sim 1\,\textrm{ms}$ is similar in magnitude to prior temperature measurements \cite{Martinis09}.  Although both hypotheses can roughly explain the data for times $\gtrsim 300\,\mu\textrm{s}$, the scenario that fits a bit better is when non-equilibrium quasiparticles are the dominant decay mechanism.

Fluctuations in the quasiparticle density cause the qubit frequency to jitter, producing dephasing \cite{dephasing}.  Using a Ramsey fringe protocol, measurements were made of the decay time $T_{2}$ at different quasiparticle injection currents, where we saw a decrease in $T_2$ at high quasiparticle densities.  After correcting for the decrease in $T_2$ from a change in $T_1$ using $1/T_2=1/2T_1+1/T_\phi$, we found that $T_\phi$ was unchanged to within experimental error.  We conclude that other dephasing mechanisms are dominant in the phase qubit.

In conclusion, we have used the theory of quasiparticle admittance to predict the qubit frequency shift and energy decay from non-equilibrium quasiparticles.  When injecting quasiparticles from a nearby junction, we find good qualitative agreement with theory.  Good quantitative agreement was observed between the relative change of qubit decay and frequency with changing quasiparticle density.  Monitoring the time dependence of qubit decay after injection should allow future experiments to positively identify if non-equilibrium quasiparticles are the limiting decay mechanism in superconducting qubits.

\begin{acknowledgments}
We thank F.\,K. Wilhelm and M.\,H. Devoret for invaluable discussions.  This
work was supported by IARPA under ARO award W911NF-09-1-0375. M.M. acknowledges support from an Elings Postdoctoral Fellowship. H.W. acknowledges
partial support by the Fundamental Research Funds for the Central Universities in China (Program No. 2010QNA3036). Devices were made at the
UC Santa Barbara Nanofabrication Facility, a part of the NSF-funded National Nanotechnology Infrastructure Network.

\end{acknowledgments}

\end{document}


\title{Supplementary material for energy decay and frequency shift \\ of a superconducting qubit from non-equilibrium quasiparticles}

\author{ M. Lenander$^1$}
\author{ H. Wang$^{1,2}$}
\author{ Radoslaw C. Bialczak$^1$}
\author{ Erik Lucero$^1$}
\author{ Matteo Mariantoni$^1$}
\author{ M. Neeley$^1$}
\author{ A. D. O'Connell$^1$}
\author{ D. Sank$^1$}
\author{ M. Weides$^1$}
\author{ J. Wenner$^1$}
\author{ T. Yamamoto$^{1,3}$}
\author{ Y. Yin$^1$}
\author{ J. Zhao$^1$}
\author{ A. N. Cleland$^1$}
\author{ John M. Martinis$^1$}

\affiliation{$^1$Department of Physics, University of California, Santa Barbara, CA 93106, USA}
\affiliation{$^2$Department of Physics, Zhejiang University, Hangzhou 310027, China}
\affiliation{$^3$Green Innovation Research Laboratories, NEC Corporation, Tsukuba, Ibaraki 305-8501, Japan}

\date{\today}

\begin{abstract}
Supplementary material is presented for the paper ``Energy decay and frequency shift of a superconducting qubit from non-equilibrium quasiparticles''.  First, we discuss quasiparticle data for a superconducting coplanar resonator.  We then document how the energy dependence of the occupation can be calculated numerically.  We calculate analytically the total quasiparticle decay rate and time dependence with both a bulk model and, numerically, one including diffusion.  We also compute the quasiparticle dependence of the gap, occupation probability, current-phase relationship, and how the frequency shift and dissipation are related.  Finally, we calculate the Josephson current for non-equilibrium quasiparticles.
\end{abstract}


\maketitle

We are interested in how non-equilibrium quasiparticles affect the properties of a Josephson junction in qubit devices.  Since we are concerned with very low temperature operation, we consider phonon temperatures sufficiently low that no quasiparticles exist due to thermal generation.  The non-equilibrium quasiparticles are generated with an unknown mechanism, and then relax their energy via the emission of phonons.  The quasiparticles typically have energy $E$ close to the gap, from which they eventually decay via recombination.  From electron-phonon physics, we know that the qusiaparticles relax to energies very near the gap, as calculated in Ref.\,\cite{Martinis09}.

The factor of 2 in the definition of $n_\textrm{qp}$ comes from integration only over positive energy, whereas excitations arise from both electron states above and below the Fermi energy.  Note this integral already contains the two possible spin states in the definition of $D(E_f)$.

\section{Quasiparticles in Coplanar resonators}

\begin{figure}[t]
\includegraphics[scale=1.0]{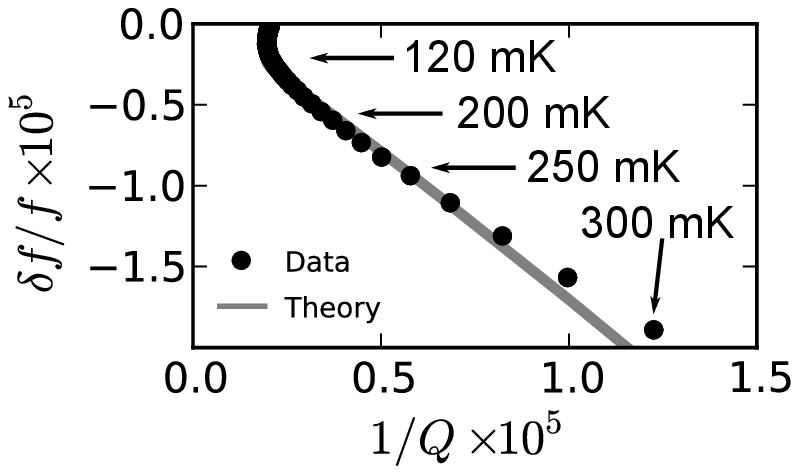}
\caption{\label{fig:resonator} Parametric plot of fractional frequency shift $\delta f/f$ versus dissipation $1/Q$, for various quasiparticle occupations $n_\textrm{qp}$ changed by varying the the sample temperature.  The device is a aluminum coplanar resonator fabricated on sapphire.  The slope of the data is in good agreement with predictions (gray line) at temperatures above 250 mK.  Deviations at low temperature are believed to come from the two-level states.  }
\end{figure}

The relationship between quasiparticle damping and frequency shift may also be tested in superconducting resonators.  Here microwave transmission is measured to extract the resonance frequency $f$ and the quality factor $Q$, with the quasiparticle density changed by simply increasing the temperature \cite{oconnell08}.  As shown in Fig.\,\ref{fig:resonator}, we find that with increasing quasiparticle density, dissipation increases and the resonance frequency decreases, in a similar manner as for junctions.

To compare with theory, we use solutions to the Mattis-Bardeen conductivity that is valid for the regime $kT \sim \hbar \omega \ll \Delta $ \cite{MB, MBints}.  These results can be expressed in terms of an admittance function discussed in the main article, but with a modification of the the 1+i term
\begin{align}
\frac{Y_j(\omega)}{1/R_n}  & = (d_R+id_I) \sqrt{2}\left(\frac{\Delta}{\hbar\omega}\right)^{3/2}
\frac{n_{\mathrm{qp}}}{n_{\mathrm{cp}}}-i\pi\frac{\Delta}{\hbar\omega} \ ,\\
d_R &= 2\sqrt{2x/\pi} \, \sinh(x) K_0(x) \ ,\\
d_I &= \sqrt{2\pi x} \, \exp(-x) I_0(x) \ ,
\end{align}
where $x=\hbar\omega/2kT$.  The theoretical prediction, indicated by the gray line in Fig.\,\ref{fig:resonator}, is in reasonable agreement with the data at high temperatures.  The experimental slope at temperatures above 200\,mK is 0.77 times that given by theory.  The deviation is largest below about 120\,mK, and believed to arise from two-level states that are not included in this model.

\section{Numerical Solution of quasiparticle recombination}

For numerical computations of non-equilibrium quasiparticle density from relaxation and recombination \cite{Martinis09}, the integrals over energy have to be put into discrete form.  For a binning size given by $d\epsilon$ in energy, we define the number of excitations in bin $i$ as
\begin{align}
n_i &\equiv d\epsilon \rho(\epsilon_i) f(\epsilon_i) \ .
\end{align}
Using this definition, the total quasiparticle density is
\begin{align}
\frac{n_\textrm{qp}}{n_\textrm{cp}} = \frac{2}{\Delta} \sum_i n_i
\end{align}

The scattering and recombination rates of Eqs.\,(6) and (7) of Ref.\,\cite{Martinis09} can then be expressed in a discrete form
\begin{align}
\Gamma_{i \rightarrow j}^s & = \sum_{j}
\Big[ \frac{(\epsilon_i-\epsilon_j)^2}{\tau_0 (kT_c)^3}
\big(1-\frac{\Delta^2}{\epsilon_i\epsilon_j}\big) N_p(\epsilon_i-\epsilon_j) d\epsilon\rho(\epsilon_j) \Big] \label{gammans}\\
&\equiv \sum_j G_{ij}^s \ ,\\
\Gamma_{i,j}^r & =\sum_{j}
\Big[ \frac{(\epsilon_i+\epsilon_j)^2}{\tau_0 (kT_c)^3}
\big(1+\frac{\Delta^2}{\epsilon_i\epsilon_j}\big) N_p(\epsilon_i+\epsilon_j) \Big] d\epsilon \rho(\epsilon_j) f(\epsilon_j) \
\label{gammanr}\\
& \equiv \sum_j G_{ij}^r \ n_j \ ,
\end{align}
where the phonon occupation factor is $N_p(E)=1/|\exp(-E/kT_p)-1|$ and the bracketed terms are the $G$ factors.  We have also assumed small occupation, so that in Eq.\,(\ref{gammans}) we use $1-f\rightarrow 1$.

The coupled differential equations for the change in the excitation number are
\begin{align}
\frac{d}{dt} n_i &= G_{ji}^s n_j-\sum_j G_{ij}^s n_i
-\sum_{j} (1+\delta_{ij})G_{ij}^r n_j n_i \ ,
\end{align}
where $\delta_{ij}$ is the Kronecker delta and accounts for the annihilation of 2 quasiparticles when in the same bin ($i=j$).  With the physics expressed in matrix form, a solution can be readily solved numerically.

\section{Quasiparticle decay}

The physics of quasiparticle relaxation and recombination was discussed in Ref.\,\cite{Martinis09}.  Although the article described solutions for the non-equilibrium occupation $f(E)$ using numerical methods, quasiparticle decay physics can be understood in the case of low density where they are mostly occupied at the gap.  The electron-electron recombination rate of a single quasiparticle, starting from Eq.\,(7) of Ref.\,\cite{Martinis09}, can be well approximated using
\begin{align}
\Gamma^r \simeq &\frac{1}{\tau_0} \int_\Delta^\infty
d\epsilon' \frac{(\epsilon+\epsilon')^2}{(kT_c)^3}
(1+\frac{\Delta^2}{\epsilon\epsilon'})\rho(\epsilon') f(\epsilon')
\label{gammar.1}\\
\simeq &\frac{1}{\tau_0}
\frac{(2\Delta)^2}{(kT_c)^3}
(1+\frac{\Delta^2}{\Delta^2}) \int_\Delta^\infty d\epsilon' \rho(\epsilon') f(\epsilon')\\
= & \frac{4}{\tau_0}
(1.76)^3 \frac{n_\textrm{qp}}{D(E_F)\Delta}\\
=&\frac{21.8}{\tau_0}\frac{n_\textrm{qp}}{n_\textrm{cp}} \ ,\label{gammar.4}
\end{align}
where we have used the BCS result $\Delta/kT_c=1.76$.  Here, $D(E_f)/2$ is the single-spin density of states, and we define the Cooper pair density $n_\textrm{cp} \equiv D(E_F)\Delta$.

The time dependence of the quasiparticle density can be understood via the rate equation
\begin{align}
\frac{d}{dt}n_\textrm{qp} &= - 2 \Gamma^r n_\textrm{qp} + r_\textrm{qp} \\
\frac{d}{dt}\frac{n_\textrm{qp}}{n_\textrm{cp}} &= - \frac{43.6}{\tau_0} \Big( \frac{n_\textrm{qp}}{n_\textrm{cp}} \Big)^2 + \frac{r_\textrm{qp}}{n_\textrm{cp}} \ ,
\end{align}
where a recombination event removes 2 quasiparticles, $r_\textrm{qp}$ is the single particle quasiparticle injection rate.  The second equation is for the normalized quasiparticle density, and has a recombination rate that is proportional to $n_\textrm{qp}^2$ because of the two-body electron-electron interaction.

The equilibrium quasiparticle density is given by setting $dn_\textrm{qp}/dt=0$, yielding density and recombination rates
\begin{align}
\frac{(n_\textrm{qp})^\textrm{eq}}{n_\textrm{cp}} &=
\Big[ \frac{\tau_0}{43.6}\frac{r_\textrm{qp}}{n_\textrm{cp}} \Big]^{1/2}
= \frac{\tau_0}{43.6} (\Gamma^r)^\textrm{eq} \ ,\label{eq:neq}\\
(\Gamma^r)^\textrm{eq} &=\Big[ \frac{43.6}{\tau_0} \frac{r_\textrm{qp}}{n_\textrm{cp}} \Big]^{1/2}
= \frac{43.6}{\tau_0} \frac{(n_\textrm{qp})^\textrm{eq}}{n_\textrm{cp}} \ .
\end{align}
The first equation is close to what was found numerically in Ref.\,\cite{Martinis09}.  The second is given by the geometric mean of the normalized injection and the characteristic electron-electron interaction rates.

We compared the results of this simple calculation with numerical solutions for a range of injection rates and found excellent agreement for $n_\textrm{qp}/n_\textrm{cp} \lesssim 0.001$.  Even at large density $n_\textrm{qp}/n_\textrm{cp} = 0.1$, Eq.\,(\ref{eq:neq}) is a reasonable approximation as its prediction is only $40\,\%$ larger than that obtained via numerics.

For no injection of quasiparticles $r_\textrm{qp}=0$, the differential equation can be integrated to give
\begin{align}
\frac{n_\textrm{qp}}{n_\textrm{cp}} =  \frac{\tau_0/43.6}{t - t_0} \,\label{eqr0}
\end{align}
where $t$ is the time and $t_0$ is an integration constant, which is approximately the time at which the quasiparticles start to cool.  The solution to the differential equation for a finite injection rate is
\begin{align}
n_\textrm{qp}&=  (n_\textrm{qp})^\textrm{eq} \coth[\,(\Gamma^r)^\textrm{eq} (t-t_0) \, ] \ ,
\end{align}
where the $\coth$ term is replaced by $\tanh$ if the quasiparticle density increases with time.  At short times the term $\coth\Gamma^r t = 1/\Gamma^r t$, which then gives Eq.\,(\ref{eqr0}) and a time dependence that scales \textit{only} with the electron-phonon coupling time $\tau_0$.  There is a relatively sharp crossover to the long time behavior where the quasiparticle density $n_\textrm{qp}^\textrm{eq}$ is constant with time.  The crossover time is given by $1/(\Gamma^r)^\textrm{eq}$.

The inverse of the crossover time thus gives the equilibrium recombination rate $(\Gamma^r)^\textrm{eq}$, which is related to the density using Eq.\,(\ref{eq:neq}) and the parameter $\tau_0$.  Comparing this density with that found from the qubit decay rate allows one to determine whether quasiparticles are the limiting decay mechanism for the qubit.

\section{Quasiparticle decay with diffusion}

The analysis in the last section assumes a bulk (uniform) model where there is no diffusion of quasiparticles.  Here, we describe a numerical solution for quasiparticle decay including relaxation, recombination, and diffusion using the simple geometry of a thin superconducting disk of radius 5 mm.  We use constant quasiparticle injection throughout the disk and a large injection pulse into the center of the disk at time $t=-200\,\mu\textrm{s}$ to $t=0$.  Because diffusion depends on the quasiparticle energy, the calculation keeps track of the occupation probability for both the radius and energy variables.

\begin{figure}[t]
\includegraphics[clip,scale=0.65]{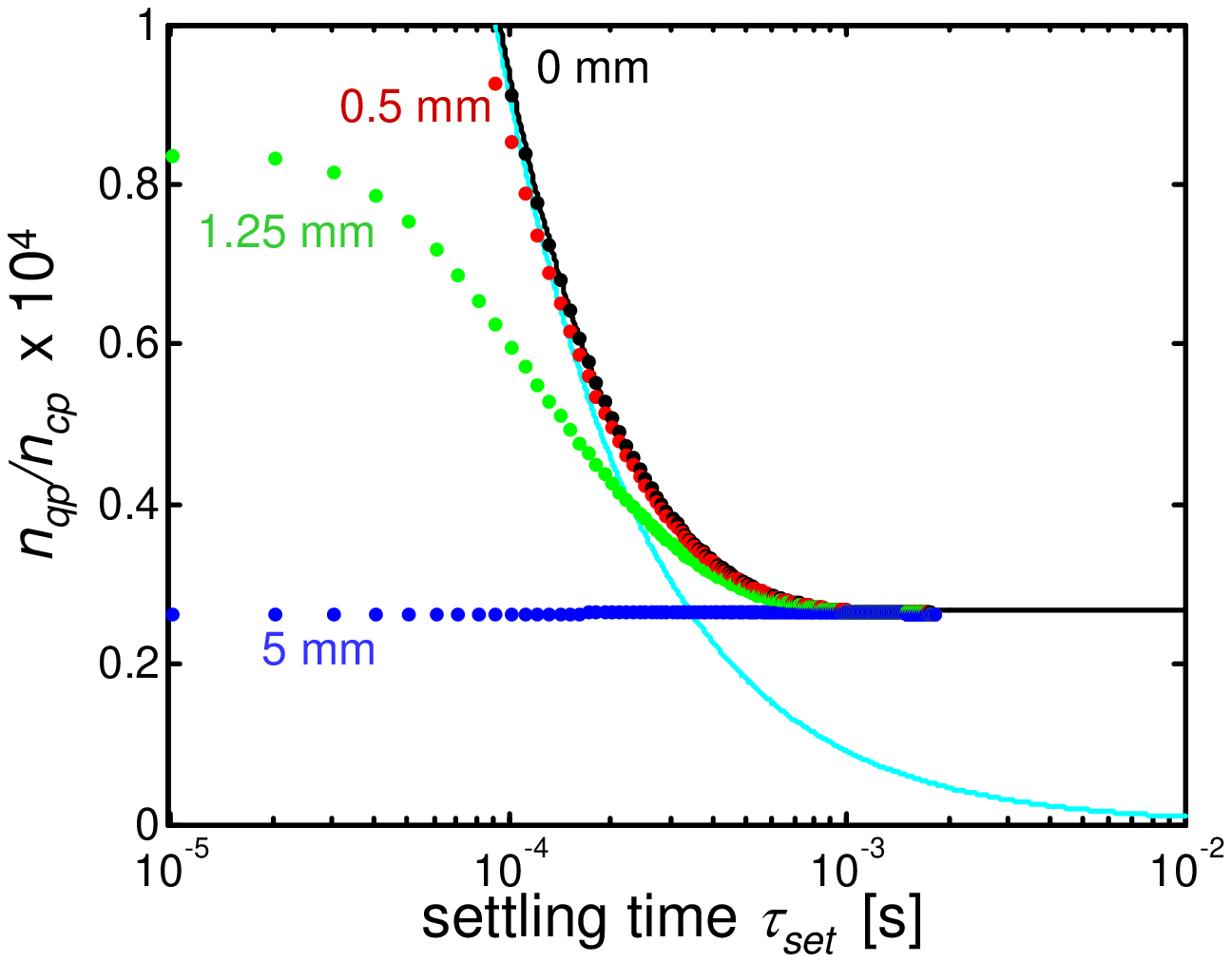}
\caption{\label{fig:qbdiffuse} Plot of quasiparticle density versus settling time for a superconducting disk of radius 5\,mm with quasiparticle injection at the center.  Points are for the simulation at four radii, r = 0 (black), 0.5\,mm (red), 1.25\,mm (green), and 5\,mm (blue).  Large changes are observed for small radii, which show a time dependence close to that predicted by the full theory of Eq.\,(9) of main article (black line) and for the zero background of Eq.\, (8) (cyan line).  At large radii much greater than the characteristic diffusion length of $\sim 1\,$mm, no change in quasiparticle density is seen.  The simulation for radius 1.25\,mm is in reasonable agreement with experimental data.   }
\end{figure}

In Fig.\,\ref{fig:qbdiffuse} we plot quasiparticle density versus settling time in a manner similar to that in the main paper, but for 4 radii.  We find differing behavior depending on the ratio of the radius with the diffusion length $\sim 1\,\textrm{mm}$, as computed for e-e diffusion in Fig.\,3 of Ref.\,\cite{Martinis09}.  For small radii, we see a dependence on time that matches closely with the bulk theory, as described in the main paper.  For a radius much larger than the diffusion length, the quasiparticle density does not change.  For the radius close to the diffusion length, we observed behavior between the two limits - a reduced peak density but a relaxation to the steady state value that has a similar time scale than for a small radius.

We note that the actual qubit device has interruptions in the ground plane due to the device geometry, so that this computation will not exactly match the experimental data.  However, the model mimics the time dependence of the data fairly well above $10\,\mu\textrm{s}$, so it is reasonable to compare to the simple bulk analysis for the behavior at long times $\gtrsim 250\,\mu\textrm{s}$.

\section{Dependence of gap on quasiparticles}

The change in the superconducting gap $\Delta$ with quasiparticles can be calculated starting from the BCS gap equation, but assuming a small non-equilibrium population $f(E)$
\begin{align}
\Delta =& D(E_f)V \int_\Delta^{\theta_D} dE\,\rho \frac{\Delta}{E} (1-2f) \ ,\\
1 = & D(E_f)V \Big(\int_\Delta^{\theta_D} \frac{dE}{\sqrt{E^2-\Delta^2}} -
\int_\Delta^{\theta_D} dE\,\rho \frac{1}{E} 2f \Big) \\
\simeq & D(E_f)V \Big( \log\frac{2\theta_D}{\Delta} - \frac{n_\textrm{qp}}{D(E_f)\Delta}\Big)
\,
\end{align}
where $V$ is the attraction potential and $\theta_D$ is the Debye energy.
Solving for the gap, one finds
\begin{align}
\Delta=&2\theta_D\exp\Big(-\frac{1}{D(E_F)V}-\frac{n_\textrm{qp}}{D(E_F)\Delta}\
\Big) \\
=&\Delta_0\exp\Big(-\frac{n_\textrm{qp}}{D(E_F)\Delta}\Big) \\
\simeq & \Delta_0\Big(1-\frac{n_\textrm{qp}}{n_\textrm{cp}}\Big)\ \label{eqBCS},
\end{align}
where $\Delta_0$ is the normal expression for the BCS gap with no quasiparticles.

\section{Numerical determination of occupation parameter}

\begin{figure}[t]
\includegraphics{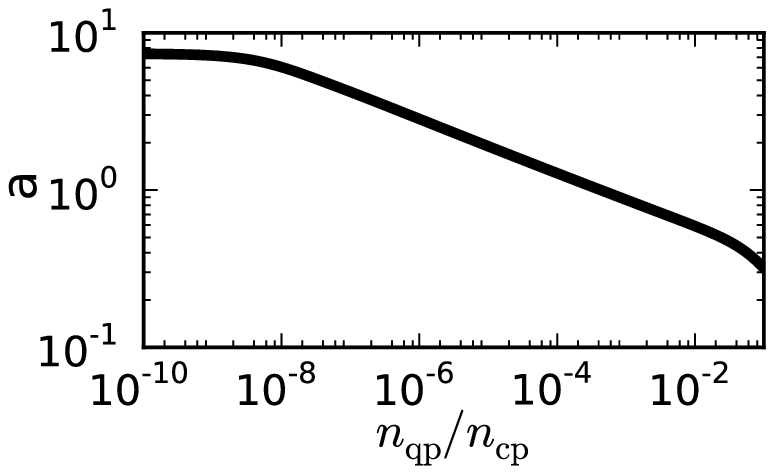}
\caption{\label{fig:aplot} Plot of $a=f(\Delta)/(n_\textrm{qp}/n_\textrm{cp})$ versus $n_\textrm{qp}/n_\textrm{cp}$, found from numerical simulation. Based on the values of $n_\mathrm{qp}/n_\mathrm{cp}$ in Fig.\,2b in the paper, we find $a \simeq 1.2$. For a wide range of injection rates, a can be well approximated by the power law $a \simeq 0.12\,(n_\textrm{qp}/n_\textrm{cp})^{-0.173}$. }
\end{figure}

To determine the effect of non-equilibrium quasiparticles, both the quasiparticle density $n_\textrm{qp}$ and the occupation probability at the gap $f(\Delta)$ must be calculated.  We plot in Fig.\,\ref{fig:aplot} the quantity $a=f(\Delta)/(n_\textrm{qp}/n_\textrm{cp})$ versus $n_\textrm{qp}/n_\textrm{cp}$ obtained from numerical computations for wide range of injection rates.  We find the results are well approximated by a line on the log-log plot, implying that the dependence can be well approximated by the power-law formula $a \simeq 0.12\,(n_\textrm{qp}/n_\textrm{cp})^{-0.173}$.

\section{Current-Phase relationship with quasiparticles}

The current-phase relationship from Josephson tunneling is given by
\begin{align}
I( \phi ) = I_0 \sin\phi \Big[ 1-\frac{n_\textrm{qp}}{n_\textrm{cp}} \Big] \big[ 1-2f(\Delta) \big] \ ,
\end{align}
where $I_0=\pi\Delta_0/2eR_n$ and the dependence of $\Delta$ on quasiparticles is now explicitly shown.  To first order, the fractional change in critical current is
\begin{align}
\frac{\delta I_0}{I_0} = -(1+2a)\frac{n_\textrm{qp}}{n_\textrm{cp}} \ .
\label{eqdeltaI0}
\end{align}

An interesting question is whether the quasiparticle tunneling terms should also be included in the current-phase relation.  For the junction current to only be a function of phase, it must arise for a purely inductive component of the junction admittance, which corresponds to terms with a frequency dependence that scales as $1/i\omega$.  Tunneling of free quasiparticles should not be included since it has an additional frequency dependence   $(n_\textrm{qp}/n_\textrm{cp})\sqrt{\Delta/\hbar\omega}$.  The Andreev bound states of the quasiparticles have an inductance, so the AC Josephson relation can then be used to find the current
\begin{align}
I_{ABS}(\phi) &= -\frac{\Phi_0}{2\pi}\int_0^\phi \textrm{Im}\{\omega Y_{ABS}(\phi)\} \ d\phi \label{eqiphiint}\\
&= I_0 [\phi+\sin\phi]f(\Delta) \ .
\end{align}
Note that this term increases the critical current, and if one replaces $\phi \rightarrow \sin\phi$ it cancels out the decrease coming from Josephson tunneling.  Since many experiments have shown the temperature dependence of the current-phase relation is given by only Josephson tunneling, we do not use this Andreev bound state term in our calculations.  In addition, our data is not consistent with including this term since it has the effect of reducing $2a$ in Eq.\,(\ref{eqdeltaI0}) to a value below $a$.

\section{Relating Dissipation and the frequency change}

Since both dissipation and the fractional critical-current change are proportional to $n_\textrm{qp}$, the magnitude of these effects are related.
The fractional change in the qubit resonance frequency $E_{10}/h$ can be calculated knowing that its dominant scaling is $E_{10}/h \propto (I_0-I)^{1/4}$, where $I$ is the qubit bias current, giving
\begin{align}
\frac{\delta (E_{10}/h)}{E_{10}/h} &= \frac{1}{4}\frac{\delta(I_0-I)}{I_0-I} \\
&= \frac{1}{4}\frac{I_0}{I_0-I}\frac{\delta I_0}{I_0} \\
&=-\frac{1+2a}{4}\frac{I_0}{I_0-I}\ \frac{n_\textrm{qp}}{n_\textrm{cp}} \ .\label{eqdf}
\end{align}

Quasiparticle dissipation can be likewise written in terms of the quasiparticle density, provided we first re-express the capacitance $C$ into qubit parameters.  Using the Josephson inductance $L_{J0}=\Phi_0/2\pi I_0$, the qubit resonance frequency is given by
\begin{align}
\frac{E_{10}}{h} \simeq \frac{1}{2\pi}\frac{1}{\sqrt{L_{J0}C}} \ [2(I_0-I)/I_0]^{1/4} \ \ .
\end{align}
We thus calculate the decay rate of the qubit
\begin{align}
\Gamma_1 &\simeq \frac{1+\cos\phi}{\sqrt{2}} \frac{2e I_0 }{\pi \Delta C}  \Big( \frac{\Delta}{E_{10}} \Big) ^{3/2} \frac{n_\textrm{qp}}{n_\textrm{cp}} \label{eqG1.1} \\
&= \frac{1+\cos\phi}{\sqrt{2}} \frac{h }{2\pi^2\Delta}\frac{1}{L_{J0}C}  \Big( \frac{\Delta}{E_{10}} \Big) ^{3/2} \frac{n_\textrm{qp}}{n_\textrm{cp}} \label{eqG1.2} \\
&\simeq \frac{1+\cos\phi}{\sqrt{2}} \frac{h}{2\pi^2 \Delta} \frac{(2\pi E_{10}/h)^2}{\sqrt{2(I_0-I)/I_0}} \Big( \frac{\Delta}{E_{10}} \Big) ^{3/2} \frac{n_\textrm{qp}}{n_\textrm{cp}} \\
&= (1+\cos\phi) \frac{E_{10}}{h} \Big( \frac{I_0}{I_0-I} \Big)^{1/2}
\Big( \frac{\Delta}{E_{10}} \Big) ^{1/2}
\frac{n_\textrm{qp}}{n_\textrm{cp}} \ .
\label{eqG1.3}
\end{align}

By taking the ratio of Eqs.\,(\ref{eqdf}) and (\ref{eqG1.3}), the quasiparticle densities cancel out, and we can relate the dissipation to the frequency shift
\begin{align}
\frac{\delta (E_{10}/h)}{\Gamma_1} &=-\frac{1}{4} \frac{1+2a}{1+\cos\phi} \,\Big(\frac{I_0}{I_0-I}\Big)^{1/2} \Big(\frac{E_{10}}{\Delta}\Big)^{1/2} \ .
\end{align}

\section{Josephson effect for arbitrary quasiparticle occupation}

Here we calculate the effect of a non-equilibrium population of quasiparticle states on Josephson tunneling, as appropriate for qubit devices.  The current proportional to $\cos\delta$ is also evaluated for the case of gaps that are not equal.  The results are readily obtained using standard second-order perturbation theory and simple integration of intermediate formulas.

The work expands on Ref.\,\cite{Martinis}, which calculated the Josephson effect at zero temperature.  In the paper, the section on Josephson tunneling is the starting point of this calculation.

The Josephson effect is derived by calculating the second-order change in energy to a superconducting state from a tunnel junction.  The tunneling Hamiltonian in second-order perturbation theory is given by
\begin{equation}
H_{T}^{(2)}=\sum_{i} H_{T}\frac{1}{\epsilon _{i}}H_{T}\text{ ,}
\end{equation}%
where $\epsilon _{i}$ is the energy of the intermediate state $i$. \ Because
the terms in $H_{T}$ have both $\gamma ^{\dagger }$ and $\gamma $
operators, the second-order Hamiltonian has terms that transfers charge across the junction but does not change the superconducting state, thus giving a change in the energy of the state.  This differs from first-order tunneling theory, which produces current only through the real creation of quasiparticles.

Because $H_{T}$ has terms that transfer charge in both directions, $%
H_{T}H_{T}$ will produce terms which transfer two electrons to the right,
two to the left, and with no net transfer. With no transfer, a
calculation of the second-order energy gives a constant value, which has no
physical effect.  We first calculate terms for the transfer of two electrons to the right from $(%
\overrightarrow{H}_{T+}+\overrightarrow{H}_{T-})(\overrightarrow{H}_{T+}+%
\overrightarrow{H}_{T-})$.  Nonzero expectation values are obtained for only two out of the four terms, as given by
\begin{align}
\overrightarrow{H_{T}^{(2)}} &= \sum_{i} \frac{
\overrightarrow{H}_{T+}\overrightarrow{H}_{T-}+
\overrightarrow{H}_{T-} \overrightarrow{H}_{T+} }
{ \epsilon_{i} } \\
&= \sum_{i}|t|^2 \frac{
(c_L^{} c_R^\dagger)( c_{-L}^{} c_{-R}^\dagger) +
(c_{-L}^{} c_{-R}^\dagger)(c_L^{} c_R^\dagger) }
{ \epsilon_{i} } \\
&= \sum_{i}|t|^2 \frac{
(c_L^{} c_{-L}^{}) (c_R^\dagger c_{-R}^\dagger)  +
(c_{-L}^{} c_L^{}) (c_{-R}^\dagger c_R^\dagger) }
{ \epsilon_{i} } \label{HT2r}
\end{align}

The pairs of electron creation and annihilation operators can be computed, giving
\begin{align}
c_k^{} c_{-k}^{} &=
(u \gamma_0^{}+ve^{i\phi}\gamma_1^\dagger)
(u \gamma_1^{}-ve^{i\phi}\gamma_0^\dagger) \\
&\rightarrow u v e^{i\phi} (-\gamma_0^{} \gamma_0^\dagger
+ \gamma_1^\dagger\gamma_1^{}) \ \ , \\
c_{-k}^{} c_{k}^{} &\rightarrow
u v e^{i\phi} (-\gamma_0^\dagger \gamma_0^{}
+ \gamma_1^{}\gamma_1^\dagger) \ \ , \\
c_k^\dagger c_{-k}^\dagger &=
(u \gamma_0^\dagger+ve^{-i\phi}\gamma_1^{})
(u \gamma_1^\dagger-ve^{-i\phi}\gamma_0^{}) \\
&\rightarrow u v e^{-i\phi} (-\gamma_0^\dagger \gamma_0^{}
+ \gamma_1^{}\gamma_1^\dagger) \ \ , \\
c_{-k}^\dagger c_{k}^\dagger &\rightarrow
u v e^{-i\phi} (-\gamma_0^{} \gamma_0^\dagger
+ \gamma_1^\dagger\gamma_1^{}) \ \ ,
\end{align}
where we have only included pairs of quasiparticle operators $\gamma$ that leaves the superconducting state unchanged, as needed for a calculation of the energy change from tunneling.  Inserting these operators into Eq.\,(\ref{HT2r}) and defining the phase difference $\delta = \phi_L - \phi_R$, we find
\begin{widetext}
\begin{align}
\overrightarrow{H_{T}^{(2)}} =
\sum_{i} |t|^2 e^{i\delta} (u_L v_L)(u_R v_R ) &
\frac{
(-\gamma_{L0}^{} \gamma_{L0}^\dagger+ \gamma_{L1}^\dagger\gamma_{L1}^{})
(-\gamma_{R0}^\dagger \gamma_{R0}^{}+ \gamma_{R1}^{}\gamma_{R1}^\dagger) +
(-\gamma_{L0}^\dagger \gamma_{L0}^{}+ \gamma_{L1}^{}\gamma_{L1}^\dagger)
(-\gamma_{R0}^{} \gamma_{R0}^\dagger+ \gamma_{R1}^\dagger\gamma_{R1}^{}) }
{ \epsilon_{i} }
\end{align}
\begin{align}
=
\sum_{i} |t|^2 e^{i\delta} (u_L v_L)(u_R v_R ) &\Big[
\frac{
-\gamma_{L0}^{} \gamma_{L0}^\dagger \gamma_{R1}^{}\gamma_{R1}^\dagger
-\gamma_{L1}^{} \gamma_{L1}^\dagger \gamma_{R0}^{} \gamma_{R0}^\dagger
}{E_L+E_R}
+\frac{
-\gamma_{L1}^\dagger \gamma_{L1}^{} \gamma_{R0}^\dagger \gamma_{R0}^{}
-\gamma_{L0}^\dagger \gamma_{L0}^{} \gamma_{R1}^\dagger \gamma_{R1}^{}
}{-E_L-E_R} \nonumber \\
& +
\frac{
 \gamma_{L0}^{} \gamma_{L0}^\dagger \gamma_{R0}^\dagger \gamma_{R0}^{}
+\gamma_{L1}^{}\gamma_{L1}^\dagger  \gamma_{R1}^\dagger\gamma_{R1}^{}
}{E_L - E_R}
+\frac{
 \gamma_{L1}^\dagger\gamma_{L1}^{}  \gamma_{R1}^{}\gamma_{R1}^\dagger
+\gamma_{L0}^\dagger \gamma_{L0}^{} \gamma_{R0}^{} \gamma_{R0}^\dagger
}{-E_L + E_R}
\Big] \ \ ,\label{HT2G}
\end{align}
where we have computed the intermediate energy $\epsilon_i$ using a positive (negative) energy $E$ for the creation (annihilation) of a quasiparticle.  The quantity $u\,v = \Delta/2E$ describes the amplitude for the virtual quasiparticle to be both electron- and hole-like, which allows a net transfer of charge by two electrons.

We now change the sum to an integral over electron states according to
\begin{align}
\sum_i & \rightarrow N_{0L}
\int_{-\infty}^{\infty} d\xi_L \ \ N_{0R} \int_{-\infty}^{\infty} d\xi_R
\ \ = 2 N_{0L} \int_{\Delta_L}^\infty \rho_L \ dE_L \ \
2 N_{0R} \int_{\Delta_R}^\infty \rho_R \ dE_R \ \ ,
\end{align}
where $N_0$ is the normal density of states, and $\rho = E/\sqrt{E^2-\Delta^2}$ is the (normalized) superconducting density of states.

By describing the superconducting state with an occupation probability of quasiparticles $f=f(E)$, the quasiparticle operators for the creation then destruction of a quasiparticle is weighted by $1-f$, while the process of destruction then creation is weighted by $f$.  The tunneling Hamiltonian is then given by
\begin{align}
\overrightarrow{H_{T}^{(2)}} & =
|t|^2 e^{i\delta} N_{0L}N_{0R}
\ 2\int_{\Delta_L}^\infty \rho_L dE_L
\ 2\int_{\Delta_R}^\infty \rho_R dE_R
\ \frac{\Delta_L}{2E_L}
\ \frac{\Delta_R}{2E_R}\ 2G \ \ , \\
G &=-\frac{(1-f_L)(1-f_R)}{E_L+E_R}
-\frac{f_L f_R}{-E_L-E_R}
+\frac{ (1-f_L)f_R}{E_L - E_R + i\epsilon}
+\frac{ f_L(1-f_R)}{-E_L + E_R + i\epsilon}
\\
& =
-\frac{1-f_L-f_R+f_Lf_R}{E_L+E_R}
+\frac{f_Lf_R}{E_L+E_R}
+\frac{ f_R-f_Lf_R}{E_L - E_R + i\epsilon}
+\frac{ f_L-f_Lf_R}{-E_L + E_R + i\epsilon}
\\
& =
-\frac{1-f_L-f_R}{E_L+E_R}
+\textrm{P}\frac{ f_R-f_L}{E_L - E_R}
+ i \pi (f_L+f_R-2f_Lf_R)\delta(E_L-E_R)
\\
& =
-\frac{1}{E_L+E_R}
+\textrm{P}\frac{ 2f_RE_L-2f_LE_R}{E_L^2 - E_R^2}
+ i \pi (f_L+f_R-2f_Lf_R)\delta(E_L-E_R) \ \ ,
\end{align}
\end{widetext}
where $G$ comes from quasiparticle operators (bracket terms in Eq.\,(\ref{HT2G})) after removing a factor of 2 because of the pair of states 0 and 1.  We take $\epsilon \rightarrow 0+$, and the integration over the zero of energy in the denominator is performed using $1/(x+i\epsilon) = \textrm{P}(1/x) + i\pi\delta(x)$, where P is the principle part and $\delta(x)$ is the Dirac $\delta$-function.

The total second-order Hamiltonian for the tunneling of 2 electrons in both directions is
\begin{align}
H_{T}^{(2)}&=\overrightarrow{H_{T}^{(2)}}+\overleftarrow{H_{T}^{(2)}} \\
&=\overrightarrow{H_{T}^{(2)}}+\textrm{h.c.}\\
&= 2 \textrm{Re}\{ \overrightarrow{H_{T}^{(2)}} \} \ \ ,
\end{align}
where h.c. is the Hermitian conjugate.

This result can be expressed in more physical terms by noting that the junction resistance can be written as
\begin{align}
\frac{1}{R_n} = \frac{4\pi e^2}{\hbar}|t|^2N_{0R}N_{0L} \ \ .
\end{align}
In addition,   The Josephson tunneling current is given by
\begin{align}
I_j=\frac{2e}{\hbar} \frac{\partial\langle H_{T}^{(2)} \rangle}{\partial \delta} \ \ .
\end{align}
Combining all of these equations, the total Josephson current is given by the integrals
\begin{widetext}
\begin{align}
I_j =
\frac{2}{\pi e R_n}
\Big\{ &\sin\delta \ \textrm{P} \int_{\Delta_L}^\infty \rho_L dE_L
\int_{\Delta_R}^\infty \rho_L dE_R
\ \frac{\Delta_L}{E_L}
\ \frac{\Delta_R}{E_R}
\Big[
\frac{1}{E_L+E_R}
-2\frac{ f_RE_L-f_LE_R}{E_L^2 - E_R^2}
\Big] \nonumber\\
- \pi &\cos\delta \ \int_{\textrm{max}(\Delta_L,\Delta_R)}^\infty \rho_L \rho_R dE
\ \frac{\Delta_L\Delta_R}{E^2}
(f_L+f_R-2f_Lf_R) \Big\}\ \ ,\label{HTAB}
\end{align}
\end{widetext}
We note that the $\sin\delta$ term in Eq.\,(\ref{HTAB}) corresponds to Eq.\,(22) of the Ambegaokar-Baratoff calculation \cite{AB}.

We evaluate these integrals by first considering, without loss
of generality, that $\Delta_L < \Delta_R$.  Using
$\rho\Delta/E=\Delta/\sqrt{E^2-\Delta^2}$
and $y=E_R/\Delta_L>1$,
we compute that the temperature independent
term $1/(E_L+E_R)$ gives for integration over $E_L$
\begin{align}
I_{1L} = &\int_{\Delta_L}^\infty dE_L \frac{\Delta_L}{\sqrt{E_L^2-\Delta_L^2}}
\frac{1}{E_L+E_R} \\
=&
\int_1^\infty dx \frac{1}{\sqrt{x^2-1}}\
\frac{1}{x+y}\\
=& \frac{\textrm{arccosh}\,y}{\sqrt{y^2-1}}\ \ .
\end{align}
The remaining integration over $E_R$ gives \cite{integrate}
\begin{align}
I_{1LR} & = \int_{\Delta_R}^\infty dE_R \frac{\Delta_R}{\sqrt{E_R^2-\Delta_R^2}}
\ \frac{\Delta_L\textrm{arccosh}(E_R/\Delta_L)}{\sqrt{E_R^2-\Delta_L^2}} \\
& = \pi \frac{\Delta_L\Delta_R}{\Delta_L+\Delta_R} \textrm{EllipticK}\Big|\frac{\Delta_L-\Delta_R}{\Delta_L+\Delta_R}\Big|
\label{anderson} \\
&= \frac{\pi^2}{4} \Delta \hspace{30pt}
(\textrm{for }\Delta_L=\Delta_R=\Delta),
\end{align}
where the last equation uses $\textrm{EllipticK}(0)=\pi/2$.  From numerical integration, we have found that Eq.\,(\ref{anderson}) is only approximate for $\Delta_L \neq \Delta_R$.

For the next term in Eq.\,(\ref{HTAB}) that has the principle part of $E_L/(E_L^2-E_R^2)$, we first integrate over $E_L$.  With the assumption $\Delta_L < \Delta_R$, the integral always passes across the pole at $E_R$ giving
\begin{align}
I_{2L}&=\textrm{P}\int_{\Delta_L}^\infty dE_L
\frac{\Delta_L}{\sqrt{E_L^2-\Delta_L^2}} \frac{E_L}{E_L^2-E_R^2}
\\
&=\textrm{P}\int_{1}^\infty dx
\frac{1}{\sqrt{x^2-1}} \frac{x}{x^2-y^2}  \\
&=-\frac{1}{\sqrt{y^2-1}} \times
\left\{
  \begin{array}{c}
    \textrm{arctanh}\sqrt{\frac{y^2-1}{x^2-1} } \ \ (x>y) \\
    \textrm{arctanh}\sqrt{\frac{x^2-1}{y^2-1} } \ \ (x<y) \\
  \end{array}
\right|_1^\infty \\
&= 0 \ \ .
\end{align}
We thus find no contribution for $f_R$ in the total integral.

We next compute the $E_R$ integral for the $E_R/(E_L^2-E_R^2)$ term.  We define $w=E_L/\Delta_R$, and note that the integration over $E_R$ depends on whether $E_L$ is greater or less than $\Delta_R$
\begin{align}
I_{2R}&=\textrm{P}\int_{\Delta_R}^\infty dE_R
\frac{\Delta_R}{\sqrt{E_R^2-\Delta_R^2}} \frac{E_R}{E_L^2-E_R^2}
\\
&=\textrm{P}\int_{1}^\infty dz
\frac{1}{\sqrt{z^2-1}} \frac{z}{z^2-w^2}  \\
&=-\frac{1}{\sqrt{1-w^2}} \times
\left\{
  \begin{array}{c}
    0 \hspace{60pt} (w>1) \\
    \textrm{arctan}\sqrt{\frac{x^2-1}{1-w^2} } \ \ (w<1) \\
  \end{array}
\right|_1^\infty \\
&=-\frac{\pi}{2}\frac{1}{\sqrt{1-w^2}}\ \theta(\Delta_R-E_L)\ \ .
\end{align}
This result implies that when integrating over $E_L$,
no contribution comes from $\Delta_R$ to infinity, so the full integral is
\begin{align}
I_{2RL} = &\int_{\Delta_L}^{\Delta_R} dE_L \frac{\Delta_L}{\sqrt{E_L^2-\Delta_L^2}}
\ 2f_LI_{2R} \\
= &-\pi \int_{\Delta_L}^{\Delta_R} f_L\ dE_L  \frac{\Delta_L}{\sqrt{E_L^2-\Delta_L^2}}
\frac{\Delta_R}{\sqrt{\Delta_R^2-E_L^2}} \ \ .\label{EqI2RL}
\end{align}
In the limit where $\Delta_L$ is close to $\Delta_R$ such that $f_L$ is constant
over the region of integration, the integral can be evaluated
\begin{align}
I_{2RL} &= -\pi f_L(\Delta_L) \
\Delta_L
\textrm{EllipticK}[1-(\Delta_L/\Delta_R)^2]\\
&\simeq -\frac{\pi^2}{4}\Delta\ 2f_L(\Delta) \ \ ,
\end{align}
where in the last equation we have taken the limit
$\Delta_L \rightarrow \Delta_R = \Delta$.

For the case of equal gaps $\Delta$,
the first two integrals give a Josephson current with a $\sin\delta$ dependence
\begin{align}
I_{js} &= \frac{2}{\pi e R_n} (I_{1LR}+I_{2RL}) \sin\delta \\ &=
\frac{\pi}{2}\frac{\Delta}{e R_n} [1-2f_L(\Delta)] \sin\delta \\
&=  \frac{\pi}{2}\frac{\Delta}{e R_n}  \tanh[\Delta/2kT] \sin\delta \ \ \
\textrm{(thermal)} \ \ .
\end{align}
The last equation assumes a thermal population of quasiparticles given by $f(E)=1/[1+\exp(E/kT)]$, which yields the Ambegaokar-Baratoff formula \cite{AB} for the Josephson current.

The Josephson current can also be calculated for an arbitrary quasiparticle occupation under the assumption that the difference of the gaps $\Delta_R-\Delta_L$ is much larger than the typical width of the quasiparticle distribution.  The contribution from the principle part for the $f_R$ term is zero, as discussed before.  The contribution from $f_L$ (the lower gap side of the junction) is given by Eq.\,(\ref{EqI2RL}).  Noting that $f_L$ is peaked at $E_L = \Delta_L$, we find the current change from quasiparticles is given by
\begin{align}
I_{j2}
\simeq &-\frac{2 \sin\delta}{e R_n} \frac{\Delta_R}{\sqrt{\Delta_R^2-\Delta_L^2}}
\int_{\Delta_L}^{\Delta_R} f_L\ dE_L  \frac{\Delta_L}{\sqrt{E_L^2-\Delta_L^2}} \\
\simeq & -\frac{\sin\delta}{e R_n}  \frac{\Delta_R \Delta_L}{\sqrt{\Delta_R^2-\Delta_L^2}} \ \frac{n_{\textrm{qp}L}}{N_{0L}\Delta_L}  \ \ ,
\end{align}
Note that these equations have a contribution from quasiparticle occupation only from the left electrode, which has lower gap.  This makes sense since a more exact theory of Andreev bound states has suppression of the critical current from occupied states in the gap, which has to have energy below that of the lowest gap.

For the $\cos\delta$ term, we first note that the current diverges logarithmically for $\Delta_L \rightarrow \Delta_R$.  Assuming the quaisparticle density is constant with $f=f_l+f_R-f_Lf_R$, numerical integration gives the approximate formula
\begin{align}
I_{jc} & \simeq
-\frac{2\cos\delta}{ e R_n} \big[-0.1 + \frac{\Delta_L}{\Delta_R}-0.5\ln \big(1-\frac{\Delta_L}{\Delta_R} \big) \ \big] f
\end{align}
For the case where the gaps greatly differ, and in the limit discussed in the previous paragraph, the current is
\begin{align}
I_{jc}
& \simeq
-\frac{2\cos\delta}{ e R_n}
\rho_L(\Delta_R) \frac{\Delta_L\Delta_R}{\Delta_R^2}
\int_{\Delta_R}^\infty \rho_R \,dE \, f_R \\
& =
-\frac{\cos\delta}{ e R_n}
\frac{\Delta_L^2}{\sqrt{\Delta_R^2-\Delta_L^2}}
\ \frac{n_{\textrm{qp}R}}{N_{0R}\Delta_R} \  \ ,
\label{Icos}
\end{align}
which has a form and magnitude similar to the thermal current.  

Note that for the $\cos\delta$ current, quasiparticles contribute from the higher gap side of the junction.  This contrasts the behavior of the $\sin\delta$ current, which has contribution from the superconducting electrode with lower gap.